\documentclass[a4paper,11pt]{article}
\pdfoutput=1

\usepackage{jcappub} 
\usepackage{natbib}
\setcitestyle{square,comma,numbers,sort&compress}
\usepackage{enumerate}
\usepackage{tabularx}
\usepackage[left = 1.0in, top=1.0in,right=1.0in,bottom=1.0in,a4paper]{geometry}
\usepackage[dvipsnames]{xcolor}
\usepackage{comment}
\usepackage[caption=false]{subfig}
\usepackage{cancel}

\usepackage[section]{placeins}
\usepackage{listings}
\usepackage{amsbsy}
\newcolumntype{Y}{>{\centering\arraybackslash}X}

\definecolor{myRED}{rgb}{0.8, 0.25, 0.33}
\usepackage[normalem]{ulem}
\usepackage{dsfont}
\usepackage{pbox}
\usepackage{graphicx}
\usepackage{multirow}
\usepackage{tikz}
\usetikzlibrary{arrows}
\usepackage{enumitem} 
\usepackage{ulem}
\usepackage{soul}
\usepackage{lipsum} 
\usepackage{bbm}
\usepackage{amsmath}
\usepackage{amssymb}
\allowdisplaybreaks

\newcommand\scalemath[2]{\scalebox{#1}{\mbox{\ensuremath{\displaystyle #2}}}}
\title{\boldmath\huge  
Fermion masses and mixings in supersymmetric SO(10) with third-generation quasi-Yukawa unification }

\author[a]{Shaikh Saad$^\dagger$,}
\author[b]{Qaisar Shafi}

\emailAdd{$^\dagger$shaikh.saad@ijs.si}

\affiliation[a]{Jožef Stefan Institute, Jamova 39, P.\ O.\ Box 3000, SI-1001 Ljubljana, Slovenia}

\affiliation[b]{Bartol Research Institute, Department of Physics and Astronomy, University of Delaware, Newark, DE 19716, USA}

\abstract{
We discuss the charged and neutral fermion masses and mixings in a supersymmetric SO(10) model with a minimal Higgs sector consisting of the multiplets $10_H$, $45_H$, and $16_H + \overline{16}_H$. In addition to the renormalizable Yukawa couplings involving the Higgs 10-plet, we include non-renormalizable Yukawa couplings, which are important for reproducing with good accuracy the observed masses and mixings in the quark and lepton sectors. We identify a preferred solution which is compatible with third family quasi-Yukawa unification, namely $y_t \approx y_b \approx 0.73 y_{\tau}$ at the unification scale, with the MSSM parameter $\tan\beta \sim 58$. Acceptable solutions with lower $\tan\beta$ values are also realized in our framework, and we provide an example with $\tan \beta =10$. Based on our fits,  the masses for the three right-handed neutrinos turn out to be $\left(M_1,M_2,M_3\right)\sim \left(10^9,  8\cdot 10^{12}, 9\cdot 10^{12}\right) \mathrm{GeV}$.  
We briefly discuss the metastable and quasistable string scenarios that can be realized in this class of supersymmetric SO(10) models.
}

\makeatletter
\gdef\@fpheader{}
\makeatother
\begin{document}
\maketitle
\flushbottom

\section{Introduction}
Grand unification based on SO(10)~\cite{Georgi:1974my, Fritzsch:1974nn}, has a number of attractive features that include unification of each family in the 16-dimensional spinor representation, and the prediction of right handed neutrino~\cite{Pati:1974yy}, one per family. In addition to a topologically stable superheavy magnetic  monopole~\cite{tHooft:1974kcl,Polyakov:1974ek}, SO(10) also predicts the presence of topologically stable cosmic strings~\cite{Kibble:1982ae}, and a large variety of composite topological structures~\cite{Lazarides:2023iim,Lazarides:2023ksx}.

Supersymmetric  SO(10), under suitable assumptions, predicts third family Yukawa coupling unification~\cite{Ananthanarayan:1991xp}, as well as quasi-Yukawa unification~\cite{Gomez:2002tj,Dar:2011sj,Shafi:2023sqa}, which we discuss in this paper. More recently, supersymmetric SO(10) has been employed~\cite{Antusch:2023zjk,Antusch:2024nqg,Fu:2023mdu} to implement the so-called metastable string scenario~\cite{Buchmuller:2021mbb}, which is designed to explain the observed stochastic gravitational  background detected by the Pulsar Timing Array (PTA) experiments~\cite{NANOGrav:2023gor,EPTA:2023fyk,Reardon:2023gzh,Xu:2023wog}.

Motivated by these more recent cosmological applications of supersymmetric SO(10), our main goal here is to see how the observed fermion masses and mixings can be incorporated in this framework.
We follow Ref.~\cite{Babu:1998wi} by employing a minimal Higgs sector consisting of $10_H$, $45_H$ and $16_H$, $\overline{16}_H$. We also take advantage, as in~\cite{Babu:1998wi}, of suitable non-renormalizable Yukawa couplings in order to realize the observed fermion masses and mixings.
Our finding, it turns out, is in good agreement with the so-called quasi-Yukawa unification scenario, according to which the top and bottom quark Yukawa couplings are essentially equal at the SO(10) breaking scale $M_\mathrm{GUT}$. However, b-tau Yukawa unification at $M_\mathrm{GUT}$ is modified in order to obtain the desired mass spectrum. For this case, the Minimal Supersymmetric Standard Model  (MSSM) parameter $\tan \beta \sim m_t / m_b$. We also find acceptable solutions with lower values of tan beta and provide an example with $\tan\beta$=10.
It is encouraging that realistic supersymmetric SO(10) models are available for realizing the metastable and quasistable cosmic string scenarios that can also be tested in the ongoing and planned gravitational wave experiments. We provide an example of one such scenario.

In summary, the novel direction pursued in our present work includes the proposal to incorporate a specific non-renormalizable operator that introduces a group-theoretical factor of `9' in the tau-lepton sector, thereby enabling the realization of third-generation quasi-Yukawa unification in the form $y_t : y_b : y_\tau = 1+\zeta : 1+ \zeta : 1-9 \zeta$ (with $\zeta\simeq 0.03$). Our proposal provides a group-theoretical origin for achieving quasi-Yukawa unification for the third generation in supersymmetric scenarios where threshold corrections are negligible. To the best of our knowledge, this is the first demonstration within the SO(10) framework of a consistent fit to the full three-generation fermion masses and mixings—including both charged fermions and neutrinos—that predicts third-generation quasi-Yukawa unification, fully compatible with the current experimental data.

Finally, let us reiterate that we are primarily concerned here with the observed fermion masses and mixings, and a nuanced discussion of proton decay, taking into account both dimension five and six operators, lies beyond the scope of this paper. For a discussion of proton decay in this framework, see Refs.~\cite{Babu:1998wi} and~\cite{Babu:2010ej}.

\section{Fermion Masses in Supersymmetric SO(10)}
The Standard Model (SM) fermions of each generation are contained in a 16-dimensional irreducible representation of SO(10). An important feature of SO(10)  is that each 16-plet also includes a right-handed neutrino, which is a singlet under the SM. The right-handed neutrinos typically acquire masses close to the Grand Unified Theory (GUT) scale, which naturally leads to tiny neutrino masses, as observed in experiments. From the product decomposition
\begin{equation}
16\times 16 = 10_s+ 120_a + 126_s,
\end{equation}
the fermion masses at the renormalizable level can arise through the $10_H$, $120_H$, and $\overline{126}_H$ dimensional Higgs representations (the Yukawa sector of SO(10)  GUTs has proven to be highly predictive, a feature that has been thoroughly explored in the literature~\cite{Lazarides:1980nt,Babu:1992ia, Bajc:2001fe,Bajc:2002iw,Fukuyama:2002ch,Goh:2003sy,
Goh:2003hf,Bertolini:2004eq, Bertolini:2005qb, Babu:2005ia,Bertolini:2006pe, Bajc:2008dc,
Joshipura:2011nn,Altarelli:2013aqa,Dueck:2013gca, Fukuyama:2015kra, Babu:2016cri, Babu:2016bmy, Babu:2018tfi, Babu:2018qca, Ohlsson:2018qpt, Ohlsson:2019sja,Babu:2020tnf,Mummidi:2021anm,Saad:2022mzu,Haba:2023dvo, Kaladharan:2023zbr,Babu:2024ahk,Fong:2025aya}). Since, the latter two representations are of higher dimensionality, in this paper we prefer   to utilize 10-dimensional representation, along with the 45-dimensional and $16+\overline{16}$ representations.

\subsection{Quasi Yukawa Unification}
\begin{figure}[t!]
\centering
\includegraphics[width=0.45\textwidth]{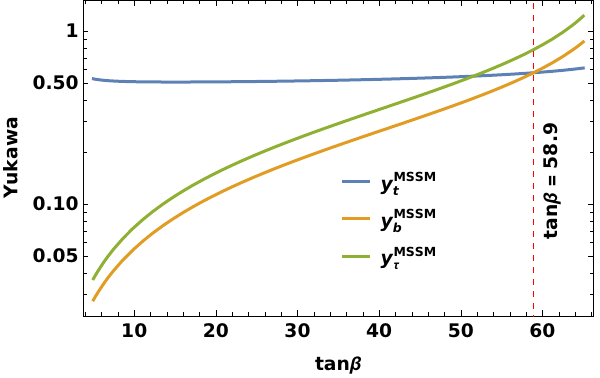}
\includegraphics[width=0.47\textwidth]{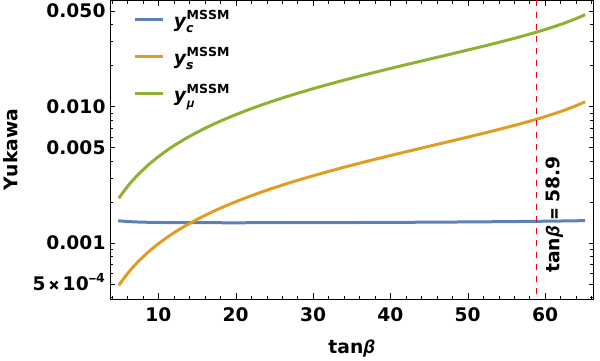}
\caption{ Left (right) panel: Third (second) generation Yukawa couplings at the GUT scale as a function of $\tan\beta$. Data is taken from Ref.~\cite{Antusch:2013jca} for the case of zero  threshold corrections from the supersymmetric partners.  }\label{fig1}
\end{figure}

Let us first concentrate on the masses of the third generation fermions and consider the renormalizable Yukawa coupling,
\begin{align}
W_{\text{Yukawa}}\supset  h_{33}  16_316_310_H. \label{eq:01}
\end{align}
Here the subscript $i$ in $16_i$ denotes the generation index. 
The $10_H$ representation contains both the up-type ($H_u$) and down-type ($H_d$) MSSM Higgs doublets, and the electroweak (EW) symmetry breaking gives rise to fermion masses. This setup automatically leads to the prediction that at the GUT scale~\cite{Ananthanarayan:1991xp}
\begin{align}
y_t^\mathrm{MSSM}=     y_b^\mathrm{MSSM} = y_\tau^\mathrm{MSSM}. 
\end{align}
The experimentally measured values, however, do not lead to an exact $t-b-\tau$ unification at the GUT scale, which is demonstrated in Fig.~\ref{fig1} (left panel). This plot is made from the data provided in Ref.~\cite{Antusch:2013jca}, which employs the relevant two-loop MSSM renormalization group equations (RGEs)  using the low scale values of the measured quantities.

The left panel of Fig.~\ref{fig1} suggests that for a large value for $v_u/v_d$, specifically, $\tan\beta\sim 58.9$, top-bottom Yukawa unification can be realized. For this value of  $\tan\beta$,
\begin{align}
\frac{ y_t^\mathrm{MSSM}}{ y_b^\mathrm{MSSM}} =1  \quad \mathrm{and}\quad   \frac{ y_\tau^\mathrm{MSSM}}{ y_b^\mathrm{MSSM}} \approx 1.36, \label{eq:02}
\end{align}
where the latter ratio is independent of $\tan\beta$. This suggests that an additional contribution to the third-generation fermion masses is necessary to account for the $b$–$\tau$ splitting. We propose to achieve this via the following higher-dimensional operator:
\begin{align}
W_{\text{Yukawa}}\supset  \frac{s_{33}}{\Lambda^2}  16_316_310_H45_H^2,   \label{eq:03}
\end{align}
where 
$\Lambda$ represents a cutoff scale (we will later provide an estimate for this $\Lambda$). 
The non-renormalizable term in Eq.~\eqref{eq:03} can provide a relative Clebsch-Gordan (CG) coefficient of $(-3)^2$ for the tau-lepton (for a systemic analysis of CG factors in SO(10) setup, see Refs.~\cite{Anderson:1993fe,Antusch:2019avd}). Interestingly, this somewhat large CG factor of $+9$ can overcome the $M_\mathrm{GUT}^2/\Lambda^2$ suppression in the above operator and provide the desired tau-bottom Yukawa coupling ratio as quoted above. For the operator in Eq.~\eqref{eq:03},  we  may assume that it appears from  integrating out vectorlike $16+\overline{16}$  fermions, which fixes the types of contractions we allow.  Note that a lower dimensional operator, namely $16_316_310_H45_H$, cannot break the equality $y_b=y_\tau$, since $10 \times 45 \supset 10 + 120$, with the
effective $120_H$ not coupling to $16_3 16_3$ owing to flavor antisymmetry.

The desired CG coefficient can be obtained if the adjoint Higgs $45_H$ gets a vacuum expectation value (VEV) in the $B-L$ direction, namely, 
\begin{align}
&\langle 45_H\rangle= i\tau_2\otimes \mathrm{diag}(a,a,a,0,0),  
\end{align}
where $\tau_i$ are the Pauli matrices.   To break SO(10) down to the SM (in particular MSSM), we further employ a  $16_H+ \overline{16}_H$  Higgses (with VEVs $\langle 16_H\rangle, \langle\overline{16}_H\rangle = c, \overline{c}$, with equal magnitudes, i.e., $|c|=|\overline c|$) such that the following symmetry breaking is realized:
\begin{align}
SO(10)
&\xrightarrow[45_H]{a} 
SU(3)_c\times SU(2)_{L} \times SU(2)_{R}\times U(1)_{B-L} 
\\
&  \xrightarrow[16_H+ \overline{16}_H]{c, \overline{c}}
SU(3)_c\times SU(2)_L\times U(1)_Y.
\end{align} 
We consider $\langle 45_H\rangle = M_\mathrm{GUT}$ and $\langle 45_H\rangle \geq  \langle16_H\rangle, \langle\overline{16}_H\rangle$ (recall that these VEVs can, in general, be complex in the supersymmetric scenario).

\subsection{Structure of the 2–3 Sector}
Let us  consider the 2-3 sector of the charged fermion masses. Unlike the above discussion with a single generation, now, in addition to the charged fermion masses, one must also incorporate the associated 2-3 mixing angles. In simultaneously accommodating both a non-zero mass for the second generation fermion as well as 23-mixing angle, a minimal option is to have a non-zero 23-entry with a vanishing 22-entry. Since our goal is to propose an economical but viable model, we proceed along this direction.

For the 23-entry, keeping only the  renormalizable operator of the type Eq.~\eqref{eq:01}, 
\begin{align}
W_{\text{Yukawa}}\supset   h_{23} 16_216_310_H, \label{eq:10}
\end{align}
which provides a symmetric contributions, one obtains
\begin{align}
 M_D\sim \begin{pmatrix}
     0&m_{23}\\
     m_{23}&m_b
 \end{pmatrix} \quad \mathrm{and} \quad M_E\sim \begin{pmatrix}
     0&m_{23}\\
     m_{23}&m_\tau
 \end{pmatrix}, \quad m_{b,\tau}\gg m_{23}.     
\end{align}
The eigenvalues of these matrices are given by $(m_{23}^2/m_b,m_b)$ and $(m_{23}^2/m_\tau,m_\tau)$, respectively, leading to $m_\mu/m_s = m_b/m_\tau$. However, this relation badly fails as can be seen from the right panel of  Fig.~\ref{fig1},  which shows the experimental value of this ratio, which is independent of $\tan\beta$, 
\begin{align}
\frac{ y_\mu^\mathrm{MSSM}}{ y_s^\mathrm{MSSM}} \approx 4.34.  \label{eq:04}
\end{align}

To remedy this,  we, therefore, consider an additional asymmetric  contribution to the 23-entry, which arises from the following non-renormalizable term~\cite{Babu:1998wi}:
\begin{align}
W_{\text{Yukawa}}\supset  \frac{a_{23}}{\Lambda}  16_216_310_H45_H.   \label{eq:05}
\end{align}
This term provides a relative CG coefficient of $-3$ for the charged leptons since $\langle 45_H\rangle\propto$ B-L (anti-symmetric due to the VEV structure $\langle 45_H\rangle\propto$ B-L), which plays a crucial role in obtaining the above-mentioned desired ratio for the second-generation Yukawa couplings. For the operator in Eq.~\eqref{eq:05}, we  may assume that it appears from  integrating out vectorlike $16+\overline{16}$  fermions, which fixes the types of contractions we allow.

It is important to note that the operators introduced above do not distinguish between the up-quark and down-quark sectors. Subsequently, in addition to obtaining the same mass patterns in these two sectors, one would obtain a trivial  Cabibbo-Kobayashi-Maskawa (CKM) matrix. Therefore, to break this pattern and obtain a correct value for $V_{cb}$, one simple way is to include an electroweak scale VEV from $16_H$ which contributes to   fermion masses. It is important to recall that  $16_H$ contains a down-type Higgs doublet and no  up-type Higgs doublet. Following Ref.~\cite{Babu:1998wi},  an operator of the form
\begin{align}
W_{\text{Yukawa}}\supset  \frac{g_{23}}{\Lambda}  16_216_316_H16_H,   \label{eq:06}
\end{align}
therefore, provides mass to only the charged leptons and down-type quarks. Consequently, this up–down asymmetry leads to non-zero CKM mixing angles.

\subsection{Dirac 1-2–3 Sector}
In this subsection we write down the full $3\times 3$ Dirac masses including the neutrinos. In the spirit of the preceding subsection, which is guided by economical structures, to generate the first generation fermion masses as well as to obtain the 1-2 and 1-3 mixing angles, we include a set of similar operators, labeled by, $h_{12}, a_{12}, g_{12}$. However, our numerical analysis shows that, in order to obtain an accurate fit to the precisely measured quantities, it is also necessary to include a nonzero $h_{11}$ entry. Combining all this together, the Yukawa part of the superpotential takes the form
\begin{align}
   \scalemath{0.95}{ W_{\text{Yukawa}} = 
Y_{10}^{ij} \, 16_i \, 16_j \, 10_H + 
\frac{1}{\Lambda} Y_{45}^{ij} \, 16_i \, 16_j \, 10_H \, 45_H + 
\frac{1}{\Lambda} Y_{16}^{ij} \, 16_i \, 16_j \, 16_H \, 16_H 
+ \frac{1}{\Lambda^2} Y_{s}^{ij} \, 16_i \, 16_j \, 10_H \, 45_H^2.}
\end{align}
The Dirac mass matrices for the up-quark, Dirac neutrino, down-quark, and charged leptons, respectively, in the basis $f_i M_{ij}f^c_j$, can be written down as
\begin{align}
&\frac{M_U}{v\sin\beta}= Y_U^\mathrm{MSSM}= \bigg\{Y_{10}^\mathrm{sym}   + \left( \epsilon_2 
\right) \; Y_{45}^\mathrm{anti}
+
\left(
\epsilon_2^2
\right) 
\; Y_{s}^\mathrm{sym}
\bigg\},
\\
&\frac{M^D_\nu}{v\sin\beta}=  \left(Y^D_\nu\right)^\mathrm{MSSM}=  \bigg\{Y_{10}^\mathrm{sym}  +(-3) \left( \epsilon_2 
\right) \; Y_{45}^\mathrm{anti}
+
(9)\left(
 \epsilon_2^2
\right) 
\; Y_{s}^\mathrm{sym}
\bigg\},
\\
&\frac{M_D}{v\cos\beta}=  Y_D^\mathrm{MSSM}=  \cos\gamma \bigg\{Y_{10}^\mathrm{sym}  + \left( \epsilon_2 
\right) \;  Y_{45}^\mathrm{anti}
+
\left(
\epsilon_2^2
\right) 
\;Y_{s}^\mathrm{sym} 
- \epsilon_3 \tan\gamma \;Y_{16}^\mathrm{sym}
\bigg\},
\\
&\frac{M_E}{v\cos\beta}=  Y_E^\mathrm{MSSM}=  \cos\gamma \bigg\{Y_{10}^\mathrm{sym}    +(-3) \left( \epsilon_2 
\right) \; Y_{45}^\mathrm{anti}
+
(9)\left(
 \epsilon_2^2
\right) 
\;\overline Y_{s}^\mathrm{sym}
- \epsilon_3 \tan\gamma  \;Y_{16}^\mathrm{sym}
\bigg\}.
\end{align}
Here, we have defined $\epsilon_2= a/\Lambda$  and $\epsilon_3=  c/\Lambda$ (recall, $|c|=|\overline c|$).
With our ansatz, the Yukawa coupling matrices are given by 
\begin{align}
&Y_{10}^\mathrm{sym}= \begin{pmatrix}
h_{11}&h_{12}&0\\
h_{12}&0&h_{23}\\
0&h_{23}&h_{33}
\end{pmatrix},
Y_{45}^\mathrm{anti}= \begin{pmatrix}
0&a_{12}&0\\
-a_{12}&0&a_{23}\\
0&-a_{23}&0
\end{pmatrix},
Y_{45}^\mathrm{sym}= \begin{pmatrix}
0&g_{12}&0\\
g_{12}&0&g_{23}\\
0&g_{23}&0
\end{pmatrix},
Y_{s}^\mathrm{sym}= \begin{pmatrix}
0&0&0\\
0&0&0\\
0&0&s_{33}
\end{pmatrix}.
\end{align}

We assume that two Higgs doublets remain light, namely, $H_u, H_d$, here, $H_d$ is a linear combination of $10^d_H$ and $16^d_H$. Thus, following Ref.~\cite{Babu:1998wi}, we define
\begin{align}
&H_u=10_u,\quad H_d=\cos\gamma 10_d + \sin\gamma 16_d,
\\
&\tan\beta= \frac{v_u}{v_d},\quad 
\tan\gamma= \frac{\langle 10_d\rangle}{\langle 16_d\rangle},
\end{align}
and correspondingly, the EW scale VEVs are given by:
\begin{align}
v^{10}_u=v\sin\beta, \quad   
v^{10}_d=v\cos\beta\cos\gamma, \quad   
v^{16}_d=v\cos\beta\sin\gamma,
\quad v=174.104\mathrm{GeV}.
\end{align}
The mixing between the $10^d_H$ and $16^d_H$ can arise, for example, from the following terms in the superpotential of the form~\cite{Babu:1998wi} 
\begin{align}
    W\supset M_{16} 16_H\overline{16}_H + \lambda \overline{16}_H 10_H \overline{16}_H,
\end{align}
leading to $\tan\gamma= \lambda \overline{c}/M_{16}$.

The mass matrices can be written down in a more convenient form as follows: 
\begin{align}
&M_U = 
\begin{pmatrix}
r_1 & \hspace{5pt} \epsilon' +r_2& 0 \\
-\epsilon' +r_2& 0 & \hspace{7pt} \epsilon + \sigma \\
0 & \hspace{5pt} -\epsilon + \sigma & \hspace{5pt} 1+\zeta
\end{pmatrix} m_U, \quad
M_D = 
\begin{pmatrix}
r_1 & {}\epsilon' + \eta^{\prime\prime}& 0 \\
-\epsilon' + \eta^{\prime\prime}& 0 & {}\epsilon + \eta \\
0 & \hspace{5pt} -\epsilon + \eta & \hspace{9pt} 1+\zeta
\end{pmatrix} m_D, \label{eq:mass01}
\\&
M^D_\nu = 
\begin{pmatrix}
r_1 & -3\epsilon' +r_2& 0 \\
3\epsilon' +r_2& 0 & -3\epsilon + \sigma \\
0 & 3\epsilon + \sigma & 1+9\zeta
\end{pmatrix} m_U, \quad
M_E = 
\begin{pmatrix}
r_1 & -3\epsilon' + \eta^{\prime\prime}& 0 \\
3\epsilon' + \eta^{\prime\prime}& 0 & -3\epsilon + \eta \\
0 & 3\epsilon + \eta & 1+9\zeta
\end{pmatrix} m_D. \label{eq:mass02}
\end{align}
In writing these, we have defined the following quantities: 
\begin{align}
&m_U= v\sin\beta\; h_{33},\quad m_D= v\cos\beta \cos\gamma\;  h_{33}, \quad \zeta=\epsilon^2_2\;  \frac{ s_{33}}{ h_{33}},
\quad
\epsilon= \frac{ a_{23}}{ h_{33}}\epsilon_2,\quad
\sigma= \frac{ h_{23}}{ h_{33}},
\\&
\eta= \sigma -  \frac{ g_{23}}{h_{33}}\epsilon_3\tan\gamma,
\quad 
\epsilon^\prime= \frac{ a_{12}}{ h_{33}}\epsilon_2,\quad
\eta^\prime= - \frac{ g_{12}}{ h_{33}}\epsilon_3\tan\gamma, 
\quad 
r_1=\frac{ h_{11}}{ h_{33}}, \quad r_2=\frac{ h_{12}}{ h_{33}}, \quad \eta^{\prime\prime}=\eta' +r_2.
\end{align}
Note that due to the hierarchical structure of the charged fermion masses, to a very good approximation, one obtains
\begin{align}
    \frac{m_t}{m_b}\approx \frac{m_U}{m_D} = \frac{\tan\beta}{\cos\gamma}.
\end{align}

\subsection{Neutrino Masses}
The Majorana masses for the 
right-handed neutrinos arise from the non-renormalizable superpotential couplings 
\begin{align}
  W_{\text{Yukawa}} \supset \frac{1}{\Lambda} Y^{ij}_{\nu^c} 16_i16_j \overline{16}_H\overline{16}_H .
\end{align}
Combining the Dirac neutrino masses and Majorana masses of the right-handed neutrinos give rise to tiny masses to the light neutrinos via the well-known seesaw~\cite{Minkowski:1977sc,Gell-Mann:1979vob,Yanagida:1979as,Glashow:1979nm,Mohapatra:1979ia,Schechter:1980gr,Schechter:1981cv} formula
\begin{align}
\mathcal{M}_\nu=- M^D_\nu   (M_\nu^R)^{-1}    \left(M^D_\nu\right)^T , \quad  M_\nu^R=\frac{\overline c^2}{\Lambda} Y_{\nu^c}.  \label{eq:mass03}
\end{align}
The Dirac neutrino mass matrix $M^D_\nu$ can be determined fully from a fit to the charged fermion masses and mixing. 

To incorporate the observed neutrino oscillation data, we follow the approach as in  Ref.~\cite{Babu:1998wi} and introduce the right-handed mass matrix~\cite{Babu:1998wi}
\begin{align}
    &M_\nu^R=M_R \begin{pmatrix}
x&0&z\\
0&0&y\\
z&y&1
\end{pmatrix}, 
\end{align}
where, we have defined
\begin{align}
x=\frac{\left(Y_{\nu^c}\right)_{11}}{\left(Y_{\nu^c}\right)_{33}}, \quad 
z=\frac{\left(Y_{\nu^c}\right)_{12}}{\left(Y_{\nu^c}\right)_{33}}, \quad
y=\frac{\left(Y_{\nu^c}\right)_{23}}{\left(Y_{\nu^c}\right)_{33}}, \quad
M_R=\frac{\overline c^2}{\Lambda} \left(Y_{\nu^c}\right)_{33}.  
\end{align}

\section{Numerical Analysis}

\begin{table}[t!]
\centering
\footnotesize
\resizebox{0.7\textwidth}{!}{
\begin{tabular}{|c|c|c|c|}
\hline
\textbf{Observables} & \textbf{Expt. Values} & \textbf{  Fitted Values} &\textbf{Pulls} \\
\hline\hline

$y_u\sin\beta/10^{-6}$ & $2.779$ & $2.853$ & $0.086$\\  
$y_c\sin\beta/10^{-3}$ & $1.441$ & $1.582$ & $\underline{1.94}$\\ 
$y_t\sin\beta$ & $0.5721$ & $0.5679$ & $-0.144$\\ \hline

$y_d\cos\beta/10^{-5}$ & $0.6866$ & $0.5114$ & $-\underline{1.27}$\\ 
$y_s\cos\beta/10^{-4}$ & $1.359$ & $1.372$ & $0.185$\\ 
$y_b\cos\beta/10^{-2}$ & $0.9596$ & $0.9754$ & $0.33$\\ \hline

$y_e\cos\beta/10^{-6}$ & $2.795$ & $2.697$ & $-0.649$\\ 
$y_\mu\cos\beta/10^{-4}$ & $5.905$ & $5.980$ & $0.254$\\ 
$y_\tau\cos\beta/10^{-2}$ & $1.304$ & $1.309$ & $0.079$\\ \hline

$\theta_{12}^{CKM}$ & $0.2273$ & $0.2386$ & $0.994$\\ 
$\theta_{23}^{CKM}/10^{-2}$ & $3.636$ & $3.739$ & $0.565$\\ 
$\theta_{13}^{CKM}/10^{-3}$ & $3.164$ & $3.207$ & $0.270$\\ 
$\delta_{CKM}^{c}$ & $1.208$ & $1.252$ & $0.7274$\\
\hline\hline

$\Delta m^2_{21} \text{ (eV}^2)/10^{-5}$ & $7.490$ & $7.378$ & $-0.298$\\ 
$\Delta m^2_{31} \text{ (eV}^2)/10^{-3}$ & $2.535$ & $2.571$ & $0.290$\\ \hline

$\sin^2 \theta_{12}$ & $0.3075$ & $0.2992$ & $-0.541$\\ 
$\sin^2 \theta_{23}$ & $0.5596$ & $0.5662$ & $0.236$\\ 
$\sin^2 \theta_{13}$ & $0.02193$ & $0.02256$ & $0.573$\\

$\delta^{\circ}_\mathrm{PMNS}$ & - &  16.62 & -\\
\hline\hline

$\chi^2$& - & - & 8.8  \\
\hline
\end{tabular}
}
\caption{Input values of the observables at the GUT scale $M_\mathrm{GUT}=2\times 10^{16}$ GeV for a  supersymmetry scale $M_S=3$ TeV with $\tan\beta=58.5$ (corresponding to our best fit). The value of the CP-violating phase, $\delta_\mathrm{PMNS}$, shown here reflects the benchmark fit parameters. By adjusting the input phases, alternative values of $\delta_\mathrm{PMNS}$ can be achieved without compromising the fit to other observables.}  
\label{result}
\end{table}

In this section we provide a full numerical analysis of our proposed fermion masses as given in Eqs.~\eqref{eq:mass01}, \eqref{eq:mass02}, and \eqref{eq:mass03}. These mass matrices are given at the GUT scale. Therefore, the experimentally measured values of the charged fermion masses and mixings to be fitted need to be taken at the  GUT scale. For this purpose, we use the input values of the GUT scale ($M_\mathrm{GUT}=2\times 10^{16}$ GeV, as suggested by gauge coupling unification within the MSSM setup) as given in Ref.~\cite{Antusch:2013jca} corresponding to a scenario with zero threshold corrections from supersymmetric partners (moreover, for definiteness, we fix the supersymmetry breaking scale at 3 TeV).  For observables with uncertainties $< 5\%$, we have allowed a maximum error of $5\%$. Note, however, that threshold corrections from supersymmetric partners may, in some cases, introduce even larger deviations from these values.

Our numerical analysis is based on minimizing the following $\chi^2$-function:
\begin{align}
\chi^2= \sum_i \mathrm{pull}_i^2,\;\;\; \;\;\mathrm{pull}_i= \frac{T_i-O_i}{\sigma_i}.    
\end{align} 
Here,  $O_i, \sigma_i, T_i$ respectively represent the experimentally measured value, the associated error, and theoretical prediction of a quantity labeled by $i$.    Moreover, the sum over $i$ runs over all observables, namely, the six quark masses, four CKM mixing parameters that include a CP-violating Dirac phase, three charged lepton masses, two neutrino mass-squared differences, and three leptonic mixing angles (18 observables in total). Note that the CP-violating Dirac phase in the PMNS  (Pontecorvo–Maki–Nakagawa–Sakata) mixing matrix is not yet measured experimentally, and therefore, we will not fit its value. The neutrino oscillation parameters to be fitted are taken from Ref.~\cite{Esteban:2024eli,NUFIT}.

Inspecting the structure of the charged fermion mass matrices, Eqs.~\eqref{eq:mass01}-\eqref{eq:mass02}, one sees that it contains ten parameters, namely, $m_{U,D}, \zeta, \epsilon^{(\prime)}, \eta^{(\prime)}, \sigma, r_{1,2}$. The Majorana neutrino mass matrix, on the other hand, consists of four  parameters, a mass scale $M_R$, and three Yukawa ratios $x,z,y$. As far as the fermion masses and mixing angles are concerned, all parameters can initially be taken as real. Therefore, this setup consists of 14 real parameters to fit 17 observables (magnitudes).  

However, since the CKM matrix includes a CP-violating phase, we incorporate this feature by allowing two complex parameters in the 1-2 sector. These two complex Yukawa couplings are taken to be $h_{21}, a_{21}$ that correspond to complex $r_2, \epsilon^\prime$, respectively. As a result, the Dirac neutrino mass matrix contains some complex entries.  Consequently, we find that a consistent fit to the neutrino oscillation data requires these ratios ($x,z,y$) to be complex. Including these phases, a recount of the number of parameters turns out to be 14 magnitudes and 5 phases versus 18 observables (17 magnitudes and 1 CP-violating phase).

It is worth emphasizing that the most minimal and predictive SUSY SO(10) frameworks—those that do not incorporate SUSY threshold corrections—are characterized by a total of 19 input parameters. This omission is well-motivated, as threshold corrections depend sensitively on the superpartner mass spectrum, which is theoretically unconstrained and introduces additional free parameters, thereby reducing the overall predictivity of the model. A recent example of such a minimal setup can be found in Ref.~\cite{Babu:2018tfi}, where a complete and successful fit to the fermion mass spectrum yields a total \(\chi^2 \sim 4\). As demonstrated in our work, we also adopt a similarly minimal parameter set consisting of 19 inputs (14 magnitudes and 5 phases), and obtain a comparable fit with \(\chi^2 \sim 8\) (see Table~\ref{result}). While alternative approaches in the literature do include SUSY threshold corrections, they typically require additional free parameters to achieve a successful fit, making them less constrained and less predictive in comparison. It is straightforward to include SUSY threshold corrections in our fitting procedure by enlarging the parameter space of the theory, which would likely lead to a further reduction in the total $\chi^2$.

For the fit that we present first, we let the code search for the most optimal value of $\tan\beta$. Our detailed numerical analysis yields the following values for the model parameters:
\begin{align}
&m_U= 92.6983\;\mathrm{GeV},\quad m_D= 1.57466\;\mathrm{GeV}, \quad M_R= 1.01549\times 10^{13}\mathrm{GeV},
\\
&\epsilon= -0.117822,\quad
\epsilon^\prime= 0.000679196e^{1.72497 i},\quad \eta=0.171819,\quad \eta^\prime=-0.00330325,
\\
&\sigma= 0.129594, \quad \zeta= 0.0375575,
\quad r_1= -0.000104132,\quad r_2=0.000467822   e^{-1.17967 i},
\\
&x=0.000182238 e^{2.0599 i},\quad z=0.0446069 e^{0.569925 i},\quad y=0.894937 e^{-0.869456 i}.
\end{align}
The experimental values and the associated fit predictions  are summarized in Table~\ref{result}. This is a good  fit and corresponds to a total $\chi^2=8.8$. It is worth pointing out that we have verified that including additional phases in the Yukawa couplings can, in principle, improve the fit. By allowing all possible phases, our numerical routine found a better fit corresponding to $\chi^2 = 5.4$ (instead of 8.8 as presented above).  Since we have performed the fit in terms of the original parameters, for completeness, in Appendix~\ref{appendix:A}, we present the fit values of these parameters.  As can be seen from the fit parameters, for this solution ($\tan\beta= 58.499, \cos\gamma=0.993719$), we have  
\begin{align}
    \frac{m_t}{m_b}\approx \frac{m_U}{m_D} = \frac{\tan\beta}{\cos\gamma}  =58.8687,
\end{align}
and
\begin{align}
    \frac{m_\tau}{m_b}\approx \frac{1+9\zeta}{1+\zeta}\approx 1+9\zeta = 1.338.
\end{align}

For the fit presented above, we also find that (see Appendix~\ref{appendix:A})
\begin{align}
    a\simeq c \simeq M_\mathrm{GUT}=2\times 10^{16}\mathrm{GeV}, \quad \mathrm{and} \quad\Lambda \simeq 5\times M_\mathrm{GUT}= 10^{17}\mathrm{GeV}. \label{eq:100} 
\end{align}
It is worth nothing here that the value of $\Lambda$ we have estimated in Eq.~\eqref{eq:100} is close to $M_\mathrm{P}/\sqrt{N}$, where $N$ denotes the total number of degrees of freedom in this model and $M_\mathrm{P}=2.44\times 10^{18}$ GeV is the reduced Planck mass. In our theory with $10_H, 45_H, 16_H, \overline{16}_H$, and three generations of chiral superfields $16_i$, we have $N=540$  propagating  particle
species, which results in $\Lambda \simeq M_\mathrm{P}/\sqrt{N}\simeq 1.05\times 10^{17}$ GeV, consistent with Eq.~\eqref{eq:100}. For a discussion of why the ultraviolet cutoff scale is $M_\mathrm{P}/\sqrt{N}$, see Refs.~\cite{Dvali:2007hz,Dvali:2007wp}. Given a lower Planck scale, there could be significant smearing of the gauge couplings near the unification scale from operators of the type $G_{\mu\nu}G^{\mu\nu}45^2_H/\Lambda^2$ (note that a lower dimensional operator of the form $G_{\mu\nu}G^{\mu\nu}45_H/\Lambda$ vanishes due to the antisymmetric property of $45_H$).  This effect could, in general, help with the unification of gauge couplings in the one-step breaking of the GUT symmetry to MSSM, because the higher-dimensional operators are expected to generate states that may lie slightly below the GUT scale, which could otherwise spoil successful coupling unification (for details, see Ref.~\cite{Antusch:2024nqg}).

Furthermore, the fit suggests
\begin{align}
\left(Y_{\nu^c}\right)_{33}= \frac{M_R \Lambda}{\overline c^2} \simeq \frac{5 M_R}{M_\mathrm{GUT}} \simeq \frac{1}{400},     
\end{align}
and predicts the following mass spectrum of the right-handed neutrinos:
\begin{align}
    \left(M_1,M_2,M_3\right)=\left(1.84\times 10^9, 8.87\times 10^{12}, 9.32\times 10^{12}\right) \mathrm{GeV}.
\end{align}

\begin{table}[t!]
\centering
\footnotesize
\resizebox{0.7\textwidth}{!}{
\begin{tabular}{|c|c|c|c|}
\hline
\textbf{Observables} & \textbf{Expt. Values} & \textbf{  Fitted Values} &\textbf{Pulls} \\
\hline\hline

$y_u\sin\beta/10^{-6}$ & $2.715$ & $2.872$ & $0.186$\\  
$y_c\sin\beta/10^{-3}$ & $1.408$ & $1.565$ & $\underline{2.228}$\\ 
$y_t\sin\beta$ & $0.5069$ & $0.4992$ & $-0.301$\\ \hline

$y_d\cos\beta/10^{-5}$ & $0.4971$ & $0.3839$ & $-\underline{1.138}$\\ 
$y_s\cos\beta/10^{-4}$ & $0.9838$ & $0.9657$ & $-0.367$\\ 
$y_b\cos\beta/10^{-2}$ & $0.5480$ & $0.5606$ & $0.460$\\ \hline

$y_e\cos\beta/10^{-6}$ & $2.025$ & $1.825$ & $-\underline{1.83}$\\ 
$y_\mu\cos\beta/10^{-4}$ & $4.275$ & $4.363$ & $0.412$\\ 
$y_\tau\cos\beta/10^{-2}$ & $0.7300$ & $0.7429$ & $0.353$\\ \hline

$\theta_{12}^{CKM}$ & $0.2273$ & $0.2486$ & $\underline{1.870}$\\ 
$\theta_{23}^{CKM}/10^{-2}$ & $3.952$ & $4.163$ & $1.06$\\ 
$\theta_{13}^{CKM}/10^{-3}$ & $3.440$ & $3.559$ & $0.692$\\ 
$\delta_{CKM}^{c}$ & $1.208$ & $1.286$ & $\underline{1.283}$\\
\hline\hline

$\Delta m^2_{21} \text{ (eV}^2)/10^{-5}$ & $7.490$ & $7.486$ & $-0.010$\\ 
$\Delta m^2_{31} \text{ (eV}^2)/10^{-3}$ & $2.535$ & $2.551$ & $0.129$\\ \hline

$\sin^2 \theta_{12}$ & $0.3075$ & $0.3061$ & $-0.0903$\\ 
$\sin^2 \theta_{23}$ & $0.5596$ & $0.5645$ & $0.174$\\ 
$\sin^2 \theta_{13}$ & $0.02193$ & $0.02205$ & $0.107$\\

$\delta^{\circ}_\mathrm{PMNS}$ & - &  37.3 & -\\
\hline\hline

$\chi^2$& - & - & 17.14  \\
\hline
\end{tabular}
}
\caption{Fit result with $\tan\beta=10$. The value of the CP-violating phase, $\delta_\mathrm{PMNS}$, shown here reflects the benchmark fit parameters. By adjusting the input phases, alternative values of $\delta_\mathrm{PMNS}$ can be achieved without compromising the fit to other observables.  }
\label{result10}
\end{table}

It is important to point out that unification of the top and bottom Yukawa couplings with $\tan\beta\sim 58.5$ (corresponding to $\cos\gamma \sim 1$) is a natural outcome of the SO(10) framework considered in this work. However, solutions with smaller values of $\tan\beta$ (corresponding to $\cos\gamma < 1$)  can also be obtained, and we explicitly demonstrate this by fixing $\tan\beta = 10$. The fitted values of the parameters are summarized in Appendix~\ref{appendix:B}, and the corresponding predictions are presented in Table~\ref{result10}. Comparing Table~\ref{result} and Table~\ref{result10}, one can see that the total $\chi^2$ (as well as   pulls of some of the individual  observables) suggest a solution with $\tan\beta \sim 58.5$ is somewhat preferable compared to a lower value of $\tan\beta$. Although both fits presented here prefer the CP-violating phase in the leptonic sector, $\delta_\mathrm{PMNS}$, to lie in the first quadrant, the model—when additional phases are included—generally allows solutions in other quadrants as well.

\section{Metastable/Quasistable Strings in SO(10)}
In our discussion so far we have made a reasonable assumption that the Higgs multiplets $45_H$ and $16_H$, $\overline{16}_H$ acquire the same GUT scale VEVs. For this case, the SO(10) symmetry breaking only produces a superheavy magnetic monopole that carries a single unit ($2\pi/e$) of Dirac magnetic charge as well as some color magnetic charge~\cite{Daniel:1979yz}. In order to implement a metastable string scenario in SO(10) which entails a cosmic string that can be broken through quantum tunneling of a monopole-antimonopole pair, we assume the presence of an additional Higgs 45-plet (which we label as $45^\prime_H$ and assume that it does not participate in Yukawa interactions; consequently, the previously established analysis of the charged fermion spectrum remains unaffected) such that the SO(10) symmetry breaking proceeds as follows:
\begin{align}
SO(10) &\xrightarrow[]{45_H} SU(3)_c \times  SU(2)_L \times  SU(2)_R \times  U(1)_{B-L} \label{eq:ssb1} \\& \xrightarrow[]{45^\prime_H} SU(3)_c \times  SU(2)_L \times  U(1)_R \times  U(1)_{B-L} \label{eq:ssb2}\\&\xrightarrow[]{16_H+\overline{16}_H} SU(3)_c \times  SU(2)_L \times  U(1)_Y  \;.\label{eq:ssb3}  
\end{align}
Note that the three symmetry breaking scales here are suitably adjusted to implement the metastable string scenario~\cite{Antusch:2023zjk,Antusch:2024nqg}.

The second breaking in Eq.~\eqref{eq:ssb2} above is produced by the VEV of the second Higgs 45- plet and yields an $SU(2)_R$ monopole which, together with the GUT monopole from the first breaking, is inflated away. In the third breaking the gauge symmetry $U(1)_R \times  U(1)_{B-L}$ is spontaneously broken such that the $SU(2)_R$ monopole is confined by the flux tube which effectively is the metastable string. It was recently shown~\cite{Maji:2025thf} that following the electroweak breaking the confined monopoles also carry some Coulomb flux.

In a quasistable cosmic string scenario~\cite{Lazarides:2022jgr,Lazarides:2023ksx} (for interesting correlations between proton decay and the appearance of superheavy quasistable strings, see Ref.~\cite{Maji:2024tzg}), which can be implemented in the symmetry breaking described by Eqs.~\eqref{eq:ssb1}-\eqref{eq:ssb3}, the decay of cosmic strings via the quantum tunneling of monopole-antimonopole pairs is exponentially suppressed. However, in this case the monopoles and antimonopoles associated with the strings reenter the horizon, after experiencing a certain number of inflationary e-foldings, and attach themselves to the strings. The stochastic gravitational wave spectrum produced in the quasistable string scenario is in good agreement with the current PTA measurements.

\section{Conclusions}
We have described how the observed fermion masses and mixings can be accommodated in a supersymmetric SO(10) model based on a minimal Higgs sector. In addition to the renormalizable Yukawa couplings, an essential role is played by a suitable choice of non-renormalizable Yukawa couplings. A numerical analysis shows that the model has a preference for a solution with large $\tan \beta\sim 58$, which is compatible with the postulate of third family quasi-Yukawa unification. For our model this corresponds to the asymptotic relation   $y_t : y_b : y_\tau = 1+\zeta : 1+ \zeta : 1-9 \zeta$, where the coefficient 9 is a group-theoretic factor, and $\zeta$ is estimated from our fits to be close to 0.03.  The novel group-theoretical factor proposed in this work is essential for achieving compatibility with current fermion mass data in supersymmetric scenarios where threshold corrections are negligible. We also consider low $\tan \beta$ solutions and present one with $\tan \beta$= 10. The three right handed neutrino masses are estimated to be $10^9$, $8 \cdot 10^{12}$, and $9 \cdot 10^{12}$ GeV. Finally, we briefly discuss how the metastable and quasistable cosmic string scenarios can be realized in this class of realistic SO(10) models.

\subsection*{Acknowledgments}
SS acknowledges the financial support
from the Slovenian Research Agency (research core funding No. P1-0035 and N1-0321).

\appendix
\section{Fit parameters: Variable $\tan\beta$}\label{appendix:A}
In this Appendix, we provide the fit parameters for the scenario where $\tan\beta$ is kept as a free variable. During the fitting procedure, we have restricted the cutoff parameter to be in the range $a/\Lambda, c/\Lambda \leq 0.2$. Moreover, for the two benchmark fits presented in this work, we have fixed $s_{33}=0.5$. The fitted values of the observables along with their corresponding pulls are presented in Table~\ref{result}. 
\begin{align}
&
\left( \epsilon_2, \epsilon_3  \right)=  (0.199998, 0.199849),
\\
&
\left(  \tan\beta, \tan\gamma \right)= (58.499, 0.112612),
\\
&Y_{10}^\mathrm{sym}=\left(
\begin{array}{ccc}
 -0.0000554514 & 0.0000949724\, -0.000230305 i & 0 \\
 0.0000949724\, -0.000230305 i & 0 & 0.0690099 \\
 0 & 0.0690099 & 0.532509 \\
\end{array}
\right),
\\
&Y_{16}^\mathrm{sym}=\left(
\begin{array}{ccc}
 0 & 0.0781597 & 0 \\
 0.0781597 & 0 & -0.999099 \\
 0 & -0.999099 & 0 \\
\end{array}
\right), 
\\
&Y_{45}^\mathrm{anti}=\left(
\begin{array}{ccc}
 0 & -0.000277697\, +0.00178695 i & 0 \\
 0.000277697-0.00178695 i & 0 & 0.313708 \\
 0 & -0.313708 & 0 \\
\end{array}
\right).
\end{align}

\section{Fit parameters: $\tan\beta=10$}\label{appendix:B}
For demonstration purpose,  we  present here a fit for the case with $\tan\beta=10$. The fit parameters are given by
\begin{align}
&m_U= 81.276\;\mathrm{GeV},\quad m_D= 0.900165\;\mathrm{GeV}, \quad M_R= 7.38544\times 10^{12}\mathrm{GeV},
\\
&\epsilon= 0.132395,\quad
\epsilon^\prime= 0.00078606e^{1.76188 i},\quad \eta=-0.191951,\quad \eta^\prime=-0.00423545,
\\
& \sigma= -0.144225, \quad \zeta= 0.0329991,
\quad r_1= -0.000124312,\quad r_2=0.000559835   e^{-1.11871 i},
\\
&x=0.000212585 e^{-1.16471 i},\quad z=0.0514103 e^{0.497272 i},\quad y=1.21819 e^{-0.914608 i}.
\end{align}
The fitted values of the observables along with their corresponding pulls are presented in Table~\ref{result10}.

In the following, we also list the fitted values of the original parameters: 
\begin{align}
&
\left( \epsilon_2, \epsilon_3, \tan\gamma  \right)=  (0.175964, 0.0168925, 8.97346),
\\
&Y_{10}^\mathrm{sym}=\left(
\begin{array}{ccc}
 -0.0000583216 & 0.000114735\, -0.000236262 i & 0 \\
 0.000114735\, -0.000236262 i & 0 & -0.0676635 \\
 0 & -0.0676635 & 0.469153 \\
\end{array}
\right),
\\
&Y_{16}^\mathrm{sym}=\left(
\begin{array}{ccc}
 0 & 0.0131087 & 0 \\
 0.0131087 & 0 & 0.147712 \\
 0 & 0.147712 & 0 \\
\end{array}
\right),
\\
&Y_{45}^\mathrm{anti}=\left(
\begin{array}{ccc}
 0 & -0.000398046\, +0.00205764 i & 0 \\
 0.000398046-0.00205764 i & 0 & -0.352992 \\
 0 & 0.352992 & 0 \\
\end{array}
\right).
\end{align}

This solution corresponds to
\begin{align}
    \frac{m_t}{m_b}\approx \frac{m_U}{m_D} = \frac{\tan\beta}{\cos\gamma}  =90.3,
\end{align}
and
\begin{align}
    \frac{m_\tau}{m_b}\approx \frac{1+9\zeta}{1+\zeta}\approx 1+9\zeta  = 1.3.
\end{align}

Moreover, the fit predicts the following mass spectrum of the right-handed neutrinos:
\begin{align}
    \left(M_1,M_2,M_3\right)=\left(1.56\times 10^9,   8.81\times 10^{12}, 9.19\times 10^{12}\right) \mathrm{GeV}.
\end{align}

\bibliographystyle{style}
\bibliography{reference}
\end{document}